\documentclass{svjour3}                     
\smartqed  
\usepackage{graphicx}
\usepackage{bm}
\usepackage{amsmath}
\usepackage{amssymb}
\usepackage{amsbsy}
\usepackage{latexsym}
\usepackage{multirow}
\usepackage{graphics}
\usepackage{graphicx}
\usepackage{epsfig}
\usepackage{color}
\definecolor{r}{rgb}{1,0,0}
\begin{document}

\title{Scaling of conditional Lagrangian time correlation functions of velocity and pressure gradient magnitudes in isotropic turbulence
\thanks{The authors are delighted to present this paper in the context of a symposium held in celebration of Professor Stephen B. Pope's pathbreaking contributions to turbulence and combustion research. They thankfully acknowledge the financial support from the Keck Foundation (L. C.) and the National Science Foundation (ITR-0428325 and CDI-0941530).}
}


\author{Huidan Yu        \and
        Charles Meneveau
}


\institute{H. Yu, \email{hyu36@jhu.edu}           \\
           C. Meneveau, \email{meneveau@jhu.edu}
           \at
              Department of Mechanical Engineering, Institute for Data Intensive Engineering and Science, Johns Hopkins University, Baltimore, MD 21218 \\
}

\date{Received: date / Accepted: date}

\maketitle

\begin{abstract}
We study Lagrangian statistics of the magnitudes of velocity and pressure gradients in isotropic turbulence by quantifying their correlation functions and their characteristic time scales. In a recent work \cite{rf:Yu09a}, it has been found that the Lagrangian time-correlations of the velocity and pressure gradient tensor and vector elements scale with the locally-defined Kolmogorov time scale, defined from the box-averaged dissipation-rate ($\epsilon_r$) and viscosity ($\nu$), according to $\tau_{K,r}=\sqrt{\nu/\epsilon_r}$.  In this work, we study the Lagrangian time-correlations of the absolute values of velocity and pressure gradients. It has long been known that such correlations display longer memories into the inertial-range as well as possible intermittency effects. We explore the appropriate temporal scales with the aim to achieve collapse of the correlation functions. The data used in this study are sampled from the web-services accessible public turbulence database ({\texttt{http://turbulence.pha.jhu.edu}}).  The database archives a $1024^4$ (space+time) pseudo-spectral direct numerical simulation of forced isotropic turbulence with Taylor-scale Reynolds number $Re_\lambda=433$, and supports spatial differentiation and spatial/temporal interpolation inside the database.  The analysis shows that the temporal evolution of the auto-correlations of the absolute values are determined not by the local Kolmogorov time-scale but by the local eddy-turnover time scale defined as $\tau_{e,r}= r^{2/3}\epsilon_r^{-1/3}$. However, considerable scatter remains and appears to be reduced only after a further (intermittency) correction factor of the form of $(r/L)^\chi$ is introduced where $L$ is the turbulence integral scale. The exponent $\chi$ varies for different variables. The collapse of the correlation functions for absolute values is, however, less satisfactory than the collapse observed for the more rapidly decaying strain-rate tensor element correlation functions.

\end{abstract}

\section{Introduction}
\label{sec:intro}
The study of turbulence from a Lagrangian viewpoint has a long history, with the earliest works of Taylor \cite{rf:Taylor21} and Richardson \cite{rf:Richardson26} both pre-dating Kolmogorov \cite{rf:Kolmogorov41}. The Kolmogorov 1941 theory used the constancy of the globally-averaged dissipation-rate $\langle \epsilon \rangle$ across scales to deduce, among others, the scaling properties of the wavenumber spectrum of kinetic energy. The 1941 theory was extended to account for intermittency by the introduction of the so-called refined Kolmogorov similarity hypothesis (RKSH) \cite{rf:Kolmogorov62} in 1962.  In this K62 extension of the theory, conditional statistics, based on the dissipation rate averaged in some particular subregion of the flow, acquires a central role. The local dissipation rate, usually denoted by  $\epsilon_r$, is defined according to
\begin{equation}
 \epsilon_r({\bf x}) = \frac{1}{V}\int \limits_{{\cal{R}}_r({\bf x})} 2 \nu \left[S_{ij}({\bf x}')\right]^2 d^3 {\bf x}',
 \label{eq:defepsilonr}
 \end{equation}
where $V$ is the volume of the subregion ${\cal{R}}_r({\bf x})$ (e.g. a box or a sphere)  of size $r$ centered at ${\bf x}$, $\nu$ is the kinematic viscosity of the fluid, and $S_{ij}$ is the strain-rate tensor. One of the main predictions of the RKSH relates to the longitudinal velocity increment at scale $r$, defined as $\delta_r u = [u_i({\bf x} + {\bf r}) - u_i({\bf x})](r_i/r)$. The RKSH states that in the inertial range of turbulence the statistics of $\delta_r u$ depend on $r$ and $\epsilon_r$. Therefore, from dimensional analysis, various moments of $\delta_r u$ conditioned upon a fixed value of $\epsilon_r$ will scale as $\langle \delta_r u^p \vert \epsilon_r \rangle = C_p (r \epsilon_r )^{p/3}$, essentially following Kolmogorov's 1941 postulate, but locally. Anomalous scaling results from the additional global averaging and anomalous scaling behavior of moments of  $\epsilon_r$. The existing literature to validate RKSH has focused mainly on velocity increments \cite{rf:Stolovitzky92,rf:Thoroddsen,rf:Chen93,rf:Stolovitzky94,rf:Chen95,rf:Ching08} or acceleration \cite{rf:Yeung06} in which the analysis can be performed on single snapshot measurements of the turbulent flow, i.e. based on a relatively `static' point of view of the flow.

In order to examine the RKSH in more depth, one would also like to inquire about its dynamical origin, specifically its role in the time evolution of the local structure of turbulence. It is well recognized that the dynamics of turbulence is best understood in a Lagrangian frame of reference, i.e. following fluid particles. Recent years have witnessed a strong revival of interest in Lagrangian statistics in turbulence. For reviews, see \cite{rf:Pope94,rf:Yeung02}. The dynamics of turbulence following fluid particles also plays a central role in the PDF modeling framework developed over the past two decades by S. Pope, starting with his seminal 1985 paper \cite{rf:Pope85}.

Besides which `frame' to use in the description of the dynamics, it is also important to select variables of interest that convey rich information about the flow. In recent years, there has been growing attention  placed in the dynamical evolution of the velocity gradient tensor ${\bf A}$ ($A_{ij}\equiv \partial u_i / \partial x_j$). This is due to the fact that ${\bf A}$ provides rich information about the topological and statistical properties of small-scale structure in turbulence. Pioneering studies of the Lagrangian structure and stochastic modeling of ${\bf A}$ are described in Refs. \cite{rf:Girimaji90a} and \cite{rf:Girimaji90}, respectively. The Lagrangian time evolution of ${\bf A}$ can be obtained by taking gradient of the NS equation \cite{rf:Vieillefosse82}:
\begin{equation}
\frac{dA_{ij}}{dt}=-A_{ik}A_{kj}-\frac{\partial ^2p}{\partial x_i \partial x_j}+\nu \frac{\partial ^2A_{ij}}{\partial x_k \partial x_k}.
\label{eq:Aij}
\end{equation}
As usual, $d/dt$ stands for Lagrangian material derivative, $p$ is the pressure divided by the density of the fluid, and the second and third terms on the right-hand-side of this equation are the pressure Hessian tensor and viscous term respectively. Neglecting viscous effects and assuming the pressure Hessian isotropic lead to a closed formulation known as the Restricted-Euler (RE) dynamics \cite{rf:Vieillefosse82}, \cite{rf:Cantwell92}. With analytically treatable solutions for the full tensor-level, the RE system provides a fruitful starting point for small structure modeling although there exist serious deficiencies in the RE dynamics, especially since it predicts nonphysical finite-time singularities \cite{rf:Cantwell92}.  Models have been developed to mimic  the regularization features of the neglected pressure Hessian and viscous terms. Efforts include a stochastic model in which the nonlinear term is modified to yield log-normal statistics of the dissipation \cite{rf:Girimaji90}, a linear damping model for viscous term \cite{rf:Martin98}, a tetrad model \cite{rf:Chertkov99} for pressure Hessian closure, a viscous diffusion closure \cite{rf:Jeong03}, a new stochastic dynamic model, so-called Recent Fluid Deformation closure, for both viscous and pressure Hessian terms \cite{rf:Chevillard06}, and a multi-scale model which includes energy exchange between scales \cite{rf:Biferale07}. The study of the temporal auto-correlation structure of various quantities associated with  ${\bf A}$ assists in the further developments and improvements of such models.

It has recently been confirmed \cite{rf:Yu09a} that when averaging over the entire domain, auto-correlation functions of velocity gradient tensor elements decay on timescales on the order of the mean Kolmogorov turnover time scale. This time scale is computed from the globally averaged rate of dissipation and viscosity. However, when performing the analysis in different subregions of the flow, turbulence intermittency was found to lead to large spatial variability in the decay time scales.  Remarkably, excellent collapse of the auto-correlation functions is recovered when using the `local Kolmogorov time-scale' defined using the locally, rather than the globally, averaged dissipation-rate ($\tau_{K,r}\equiv \sqrt{\nu/\epsilon_r}$). This is an additional new evidence for the validity of Kolmogorov's Refined Similarity Hypothesis, but from a Lagrangian viewpoint that provides a natural frame to describe the dynamical time evolution of turbulence.

In this paper, we study Lagrangian time-correlations of scalar measures (such as magnitudes) of velocity and pressure gradients and explore whether there is further evidence for the Lagrangian RKSH for these variables. Lagrangian correlation functions of square strain- and rotation-rate have already been studied in prior work \cite{rf:Guala07,rf:Yeung07}. In examining the scaling of the magnitudes of the velocity gradient tensor and pressure gradient, the behavior of correlation functions will be shown here to be much more complex than for the tensor or vector elements themselves. Nontrivial dependencies on local-length scale $r$ and local dissipation-rate ($\epsilon_r$) are observed, and these require more detailed study.  The present paper is devoted to such a study, based on analysis of Lagrangian data.

Lagrangian data can be extracted from direct numerical simulation (DNS) of NS equations with relative ease. The first such effort traces back to Riley and Patterson \cite{rf:Riley74}. The rapid development in computing power over the past few decades has spurred vast amount of such numerical investigations at increasing Reynolds numbers. For relevant reviews, see \cite{rf:She91,rf:Pope94,rf:Yeung02,rf:Toschi09} and references therein.
A new way to exploit large databases in turbulence has been recently proposed \cite{rf:Li08}. This approach is based on web-services that allow public access to turbulence DNS databases that store not only snapshots of 3D distributions but also the entire pre-computed time history. Using this public turbulence database, here we study the Lagrangian time evolution of velocity and pressure gradients and their magnitudes in isotropic turbulence.

The remainder of this paper is organized as follows. Section \ref{sec:database} describes the public turbulence database and the numerical approach we use to perform the Lagrangian analysis. Results on the time evolution of auto-correlations of velocity and pressure gradient magnitudes are presented in Section \ref{sec:correlations}. We conclude in Section \ref{sec:concl} with a short discussion.

\section{JHU public turbulence database and particle-tracking approach}
\label{sec:database}

The DNS data of a forced isotropic turbulence archived in the JHU public database system are from a pseudo-spectral parallel computation of the forced NS equations in a $[0, 2\pi]^3$ domain, at a Taylor-microscale Reynolds number of $Re_\lambda \simeq 433$ \cite{rf:Li08}. The database contains of output on $1024^3$ spatial points and 1024 time samples (every tenth DNS time-step is stored) spanning about one large-scale eddy turnover time. The domain-wide averaged dissipation-rate ($<\epsilon>$) and the corresponding Kolmogorov time scale ($\tau_K$) are 0.092 and 0.045 respectively, in the units of the simulation. The turbulence integral scale is $L=1.376$. Some data processing functionalities such as spatial differentiation, and spatial and temporal interpolations are provided directly inside the database. This feature not only reduces data download cost but also allows users to obtain desired quantities at arbitrary locations and times. The whole database results in a 27 Terabyte storage size. The $1024^4$ space+time history of turbulence is publicly accessible through a web-service interface which serves as a bridge to connect user requests with the database nodes. Users may write and execute analysis programs using prevailing languages C, Fortran, or Matlab on their host computers such as desktops or laptops, while the programs request desired outputs from the database through GetFunctions (subroutine-like calls) over the Internet. Currently, eight GetFunctions listed in Table \ref{ta:getfunctions} for velocity and pressure along with their derivatives and force are available. With these call functions, users can retrieve quantities simultaneously for large amounts of locations and time (within the stored time frames) without expensive memory and time costs. The details of the DNS data and JHU turbulence database can be found in a previous publication \cite{rf:Li08}. The instructions and sample codes in C, Fortran, and Matlab are available at \texttt{http://turbulence.pha.jhu.edu}.

\begin{table}[htbp]
\begin{tabular}{|c|c|c|c|c|}
 \hline
 Function name & Spatial diff. & Spatial int. & Temporal int. & Outputs\\
\hline
GetVelocity & -- & NoInt, Lag4,6,8 & NoInt, PCHIP & $u_i$\\
\hline
GetVelocityAndPressure & -- & NoInt, Lag4,6,8 & NoInt, PCHIP & $u_i,p$\\
\hline
GetVelocityGradient & FD4,6,8 & NoInt, Lag4,6,8 & NoInt, PCHIP & $\frac{\partial u_i}{\partial x_j}$\\
\hline
GetPressureGradient & FD4,6,8 & NoInt, Lag4,6,8 & NoInt, PCHIP & $\frac{\partial p}{\partial x_i}$\\
\hline
GetVelocityHessian & FD4,6,8 & NoInt, Lag4,6,8 & NoInt, PCHIP & $\frac{\partial^2 u_k}{\partial x_i\partial x_j}$\\
\hline
GetPressureHessian & FD4,6,8 & NoInt, Lag4,6,8 & NoInt, PCHIP & $\frac{\partial^2 p}{\partial x_i\partial x_j}$\\
\hline
GetVelocityLaplacian & FD4,6,8 & NoInt, Lag4,6,8 & NoInt, PCHIP & $\frac{\partial^2 u_i}{\partial x_j\partial x_j}$\\
\hline
GetForce & -- & NoInt, Lag4,6,8 & NoInt, PCHIP & $f_i$\\
\hline
\end{tabular}
\caption{Subroutine-like call functions. diff: differetiation; int: interpolation; NoInt: no interpolation; FD: Centered finite difference, options for 4th-, 6th-, and 8th-order accuracies; Lag: Lagrangian polynomial interpolation, options for 4th-, 6th-, and 8th-order accuracies; PCHIP: Piecewise cubic Hermite interpolation. }
\label{ta:getfunctions}
\end{table}

We employ the particle-tracking algorithm of Ref. \cite{rf:Yeung88} to extract Lagrangian information along many particle trajectories simultaneously. Each particle is tagged and randomly assigned an initial position. Let ${\bf x}^+({\bf y},t)$ and ${\bf u}^+({\bf y},t)$ denote the position and velocity at time t of the fluid particle originating from position ${\bf y}$ at initial time $t_0$ with the superscript $+$ representing Lagrangian quantities following the fluid particle. Each particle is tracked by numerically integrating
\begin{equation}
\frac{\partial {\bf  x}^+({\bf y},t)}{\partial t}={\bf u}^+({\bf y},t)
\label{eq:particlemotionequation}
\end{equation}
where the Lagrangian velocity ${\bf u}^+({\bf y},t)$ is replaced by the Eulerian velocity ${\bf u}({\bf x},t)$ where the particle is located, namely ${\bf u}^+({\bf y},t)={\bf u}({\bf x}^+({\bf y},t),t)$.

The particle displacement between two successive time instants $t_n$ and $t_{n+1}(=t_n+\delta t)$ is obtained through an integral of Eq. (\ref{eq:particlemotionequation}) using a second-order Runge-Kutta method. At time $t_n$ for a particle located at ${\bf x}^+({\bf y},t_n)$, the predictor step yields an estimate ${\bf x}^*={\bf x}^+({\bf y},t_n) + \delta t ~{\bf u}^+({\bf y},t)$ for the destination position ${\bf x}^+({\bf y},t_{n+1})$. The corrector step then gives the particle position at $t_{n+1}$: ${\bf x}^+({\bf y},t_{n+1})={\bf x}^+({\bf y},t_n)+\delta t ~[{\bf u}^+({\bf y},t_n)+{\bf u}^+({\bf x}^*,t_{n+1})]/2$. It is proved that the time-stepping error is of order $(\delta t)^3$ over one time step \cite{rf:Yeung88}. In general, accurate spatial and time interpolations are crucial to obtain the fluid velocities while tracking particles along their trajectories. In JHU turbulence database, these operations have been built in with optional orders of accuracy. We use flags FD4Lag4, Lag8, and PHCIP (explained in the caption of Table \ref{ta:getfunctions}) for the calls to specify spatial differentiation, spatial interpolation, and time interpolation.

\section{Lagrangian time correlations of gradient magnitudes}
\label{sec:correlations}

As mentioned above, the velocity gradient tensor ${\bf A}$ has received considerable attention in recent years. It provides a rich characterization of the topological and statistical properties of the small-scale structure in turbulence  in the viscous range. The antisymmetric part of the velocity gradient tensor, i.e., $\Omega_{ij} \equiv (A_{ij}-A_{ji})/2$, is  the rate of rotation describing the vortex structure and dynamics. Whereas the symmetric part of ${\bf A}$, the strain-rate tensor, defined as $S_{ij} \equiv (A_{ij}+A_{ji})/2$, represents the strength and directions of fluid deformation rates. The dynamic evolution of ${\bf A}$ is given by Eq. (\ref{eq:Aij}). Here we study Lagrangian auto-correlations for the absolute values of velocity-derivative tensor using the magnitudes of the strain-rate tensor ${\bf S}$ and rotation-rate tensor ${\bf \Omega}$. The magnitudes are defined here using the square invariant according to $|{\bf S}|\equiv \sqrt{S_{ij}S_{ij}}$ and  $|{\bf \Omega}|\equiv \sqrt{\Omega_{ij}\Omega_{ij}}$ (we apply Einstein notation for repeating indexes unless indicated otherwise). We also study the magnitude of pressure gradient (approximately similar to the acceleration magnitude), defined as
$|\bigtriangledown p|\equiv \sqrt{\nabla p \cdot \nabla p}$.

The Lagrangian time correlation of these scalar quantities is defined as usual:
\begin{equation}
 \rho_{f}(\tau) \equiv \frac{\langle{f(t_0)f(t_0+\tau)}\rangle}
 {\sqrt{\langle{f(t_0)^2}\rangle \cdot \langle{f(t_0+\tau)^2}\rangle}},
 \label{eq:scalarcorrelationfunction}
\end{equation}
where $\tau$ is the time-lag along Lagrangian trajectories, $f$ can be $f=|{\bf S}|$, $f=|{\bf \Omega}|$, or $f=|\bigtriangledown p |$ as the case may be,
and $\langle{\cdots}\rangle$ may represent ensemble or global volume averaging for homogeneous turbulence.

In order to study effects of intermittency, which in turbulence is characterized by local regions displaying different levels of turbulence activity,  we also compute conditional correlation functions based on fluid particles that originate from various subregions of the flow domain. The subregions are characterized by the local dissipation-rate $\epsilon_r$ defined in Eq. (\ref{eq:defepsilonr}) and the length-scale $r$. With this context, the global average in Eq.(\ref{eq:scalarcorrelationfunction}) is replaced by the conditional average, i.e

\begin{equation}
 \rho_{f}(\tau) \equiv \frac{\langle{f(t_0)f(t_0+\tau)\vert \epsilon_r}\rangle}
 {\sqrt{\langle{f(t_0)^2\vert \epsilon_r}\rangle \cdot \langle{f(t_0+\tau)^2\vert \epsilon_r}\rangle}},
 \label{eq:scalarconditional}
\end{equation}

In the conditional average the initial position of particles contributing to the average at time $t_0$ are sampled from several local boxes of size $r$ that have a prescribed locally averaged dissipation-rate $\epsilon_r$. For practical reasons, a finite
range of values of $\epsilon_r$ must be considered, i.e. we use sampling in  bins of $\epsilon_r$ values.  For each bin there are several such local cubes for which their $\epsilon_r$ falls in a prescribed range.  Besides varying the bin location, we also consider four length scales, $r=34\eta_K, 68\eta_K,136\eta_K,272\eta_K$. They correspond to 16-, 32-, 64-, and 128 grid-point cubes, respectively. Each has four associated $\epsilon_r$ bins.
An additional set of 64-cube cases ($r=136\eta_K$) with five bins has been studied. This set was also considered in \cite{rf:Yu09a} for the tensor element-based correlations. The specifications of scale and bin values are listed in Table \ref{ta:localcases}.
\begin{table}[htbp]
\begin{tabular}{|c|c|c|c|c|c|}
 \hline
 cube size $r$ & bin index & particle/cube & cube/$\epsilon_r$ & $\epsilon_r$ range & nominal $\epsilon_r$ \\
 \hline
   34$\eta_K$ & 1  & 50 & 120 & 0.0056 $\backsim$ 0.0093 & 0.0076 \\
 \cline{2-6}
  & 2 & 50 & 120 & 0.017 $\backsim$ 0.020 & 0.019 \\
  \cline{2-6}
  & 3 & 50 & 120 & 0.089 $\backsim$ 0.096 & 0.092 \\
  \cline{2-6}
  & 4 & 50 & 120 & 0.14 $\backsim$ 0.16 & 0.15 \\
 \hline
  68$\eta_K$ & 1 & 100 & 60 & 0.0074 $\backsim$ 0.011 & 0.0096 \\
 \cline{2-6}
  & 2 & 100 & 60 & 0.0241 $\backsim$ 0.028 & 0.019 \\
  \cline{2-6}
  & 3 & 100 & 60 & 0.089 $\backsim$ 0.096 & 0.093 \\
  \cline{2-6}
  & 4 & 100 & 60 & 0.14 $\backsim$ 0.16 & 0.15\\
 \hline
  136$\eta_K$-I & 1 & 500 & 12 & 0.017 $\backsim$ 0.020 & 0.018\\
 \cline{2-6}
  & 2 & 500 & 12 & 0.024 $\backsim$ 0.028 & 0.026\\
  \cline{2-6}
  & 3 & 500 & 12 & 0.089 $\backsim$ 0.096\ & 0.093\\
  \cline{2-6}
  & 4 & 500 & 12 & 0.10 $\backsim$ 0.12 & 0.11\\
  \hline
  136$\eta_K$-II & 1 & 4000 & 2 & 0.033 $\backsim$ 0.041 & 0.037\\
 \cline{2-6}
  & 2 & 4000 & 3 & 0.045 $\backsim$ 0.046\ & 0.046\\
  \cline{2-6}
  & 3 & 4000 & 2 & 0.065 $\backsim$ 0.066 & 0.066\\
  \cline{2-6}
  & 4 & 4000 & 2 & 0.083 $\backsim$ 0.088\ & 0.086\\
  \cline{2-6}
  & 5 & 4000 & 3 & 0.12 $\backsim$ 0.16 & 0.14\\
 \hline
  276$\eta_K$ & 1 & 1000 & 6 & 0.030 $\backsim$ 0.037 & 0.034\\
 \cline{2-6}
  & 2 & 1000 & 6 & 0.048 $\backsim$ 0.060 & 0.052\\
  \cline{2-6}
  & 3 & 1000 & 6 & 0.081 $\backsim$ 0.10 & 0.094\\
  \cline{2-6}
  & 4 & 1000 & 6 & 0.11 $\backsim$ 0.15 & 0.12 \\
 \hline
\end{tabular}
\caption{Indication of different cases for calculation of conditional correlation functions. Shown for local subregion (cube) length size, bin index,
number of particles in each cube, number of cubes in each bin, and range and nominal of local dissipation for each bin.}
\label{ta:localcases}
\end{table}

\begin{figure}[htbp]
   \begin{center}
      \includegraphics[width=5. in] {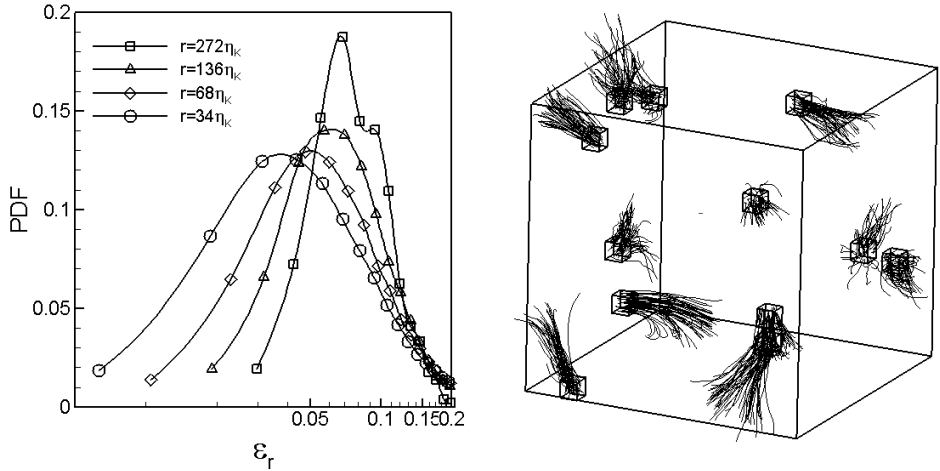}
      \caption{Left plot: PDFs of locally-averaged dissipation-rates $\epsilon_r$ with four different local length scales; Right plot: Sample particle trajectories starting from 12 randomly selected $64$-cubes characterized by local dissipation-rate $\epsilon_r$ at the initial time corresponding to case $136\eta_K-II$ in Table \ref{ta:localcases}. }
      \label{Fig_PDFtrajectory}
   \end{center}
\end{figure}

In Fig. \ref{Fig_PDFtrajectory} we show PDFs of $\epsilon_r$ for the four   length scales considered (left plot) and 12 representative $64$-cubes placed inside the $1024^3$ domain, with 50 sample fluid particle trajectories emanating from each and progressing during a time equal to 27$\tau_K$ (right plot). The required averages are taken over all the trajectories as well as over several cubes for which $\epsilon_r$ is in a bin's prescribed range.

\begin{figure}[htbp]
   \begin{center}
      \includegraphics[width=4.5 in] {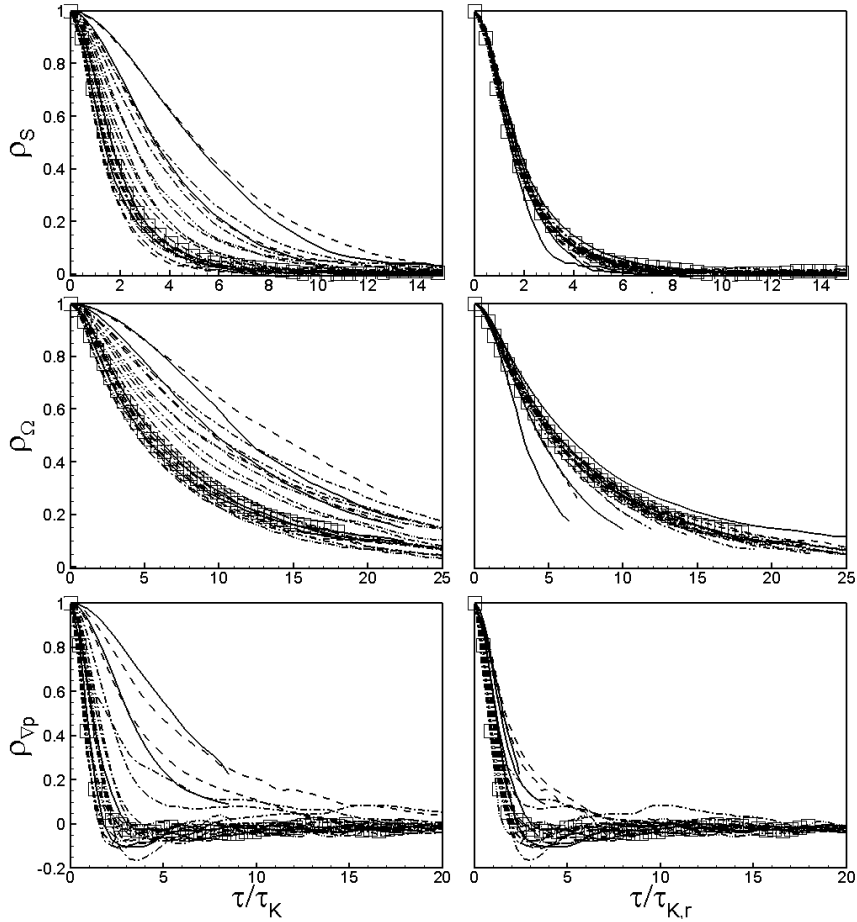}
      \caption{Lagrangian time correlations of strain- and rotation-rate tensors and pressure gradient vector for all the cases in Table \ref{ta:localcases}.  Open squares are for global average over randomly located particles in the whole domain, whereas different lines correspond to subregions of the flow characterized by different $\epsilon_r$s. Time-lag is normalized using the global Kolmogorov time scale $\tau_{K}$ (left column) and the local time-scale $\tau_{K,r}$ (right column).}
      \label{Fig_tensorcollapse}
   \end{center}
\end{figure}
Lagrangian and conditional Lagrangian auto-correlation functions for tensors ${\bf A}$, ${\bf S}$, ${\bf \Omega}$, or vector $\bigtriangledown p$ have been studied in our previous work \cite{rf:Yu09a,rf:Yu09b}. These tensor and vector time correlation functions are computed through expressions like $\langle{C_{ij}(t_0)C_{ij}(t_0+\tau)}\rangle$ and $\langle{C_{ij}(t_0)C_{ij}(t_0+\tau)\vert \epsilon_r}\rangle$ or $\langle{G_i(t_0)G_i(t_0+\tau)}\rangle$ and $\langle{G_i(t_0)G_i(t_0+\tau)\vert \epsilon_r}\rangle$ on tensor or vector element level for global and conditional correlation functions, respectively. Here we present these measurements in Fig. \ref{Fig_tensorcollapse} for three variables including all the cases listed in Table \ref{ta:localcases} (in \cite{rf:Yu09a} only results for ${\bf S}$ were shown for the same cases). The evolution time is scaled by $\tau_K$ and $\tau_{K,r}$ where
\begin{equation}
\tau_K = \sqrt{\frac{\nu}{\langle \epsilon \rangle}}, ~~~~ \tau_{K,r}= \sqrt{\frac{\nu}{\epsilon_r} }.
\end{equation}
The main observations can be summarized as follows. First, the rotation-rate displays significantly longer time memory than the strain-rate. After about 6$\tau_{K,r}$, the strain-rate's correlation is essentially zero, whereas it is still near 0.5 for the rotation-rate. We found that this trend holds true even if coherent vortex structures are excluded from the analysis. Second, the temporal auto-correlation functions scatter significantly when the time lag is scaled by $\tau_K$ (left column) but collapse well when scaled by $\tau_{K,r}$ (right column). This behavior demonstrates that the dynamics of flow variables such as velocity and pressure gradients following fluid particles depends upon the local dissipation-rate ($\epsilon_r$) rather than the global one ($\langle \epsilon \rangle$) which, as argued in \cite{rf:Yu09a}, provides new evidence for the validity of Kolmogorov's refined similarity hypothesis form a Lagrangian viewpoint.

In what follows, we study conditional Lagrangian time correlations for the absolute values of these tensors and vector.
\begin{figure}[htbp]
   \begin{center}
      \includegraphics[width=4.5 in] {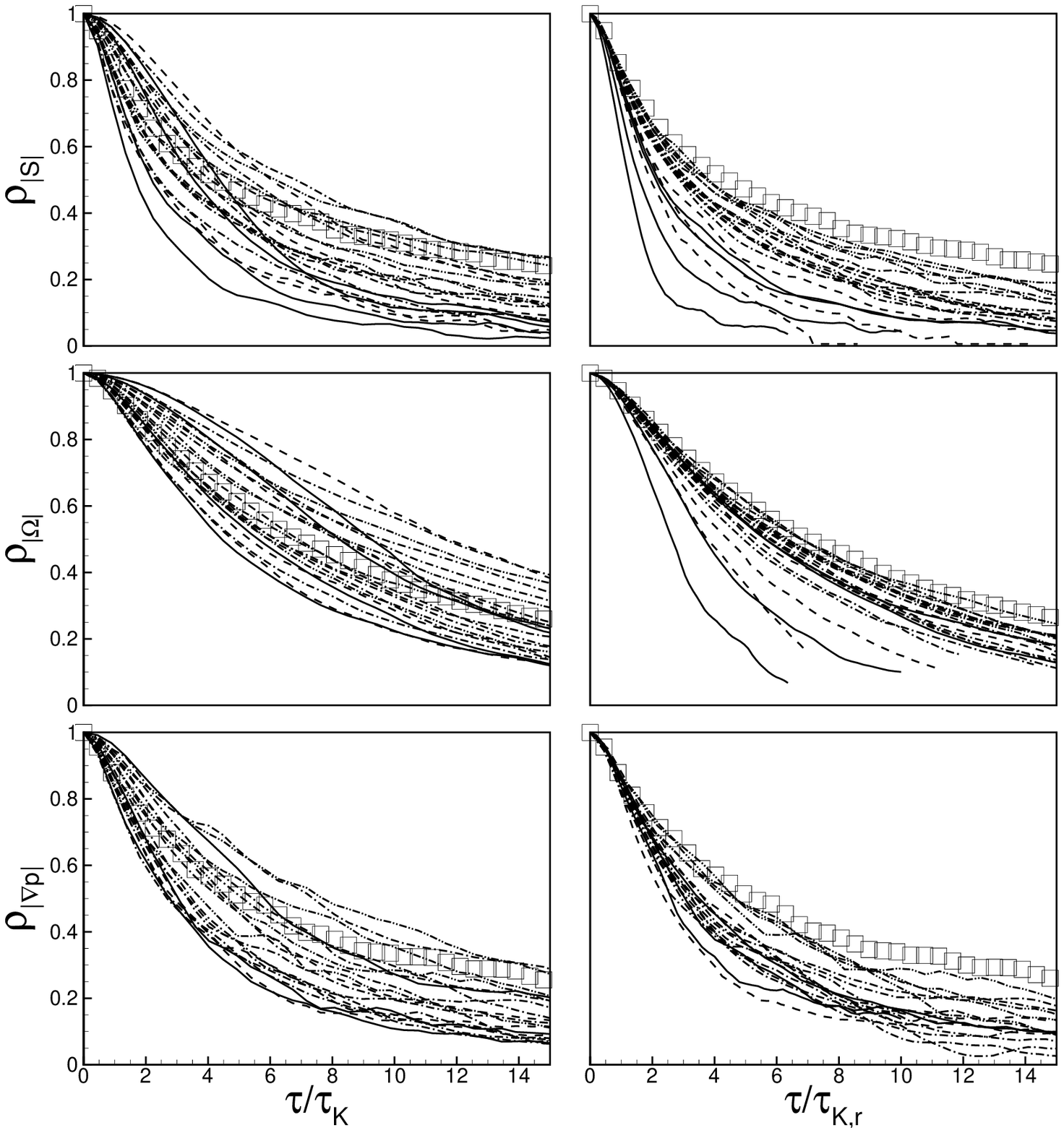}
      \caption{Auto-correlations of $|S|$ (top row), $|\Omega|$ (middle row), and $|\nabla p|$ (bottom row) vs. $\tau/\tau_K$(left column), $\tau/\tau_{K,r}$ (right column) for the cases listed in Table \ref{ta:localcases}. Open square symbols correspond to global (unconditional) averaging over entire data volume.}
      \label{Fig_scalar_tauK}
   \end{center}
\end{figure}

Figure \ref{Fig_scalar_tauK} shows the similar plots to Fig. \ref{Fig_tensorcollapse} but for the the absolute values of tensors ${\bf S}$ (top row) and ${\bf \Omega}$ (middle row) and vector $\nabla p$ (bottom row). It is quite clear that for all variables, especially for the strain-rate and pressure gradient, the correlations decay much more slowly for the magnitudes as compared to the tensor or vector elements. Similarly slow decay had been observed for the square of these
variables in \cite{rf:Guala07,rf:Yeung07}. Moreover, and unlike the tensor- or vector-based Lagrangian auto-correlations, poor collapse is seen when the time lag $\tau$ is scaled by the local Kolmogorov time $\tau_{K,r}$. Such scaling appears to work only for the viscous time-scale range near the origin of the curves ($\tau < \tau_{K,r}$). For the inertial range, the curves scatter significantly even when scaled by the local $\tau_{K,r}$.

Since there remain significant correlation scatters even after long time delays, we explore the use of other time-scales to express time. The characteristic time-scale that is believed to be relevant in the inertial range is the eddy-turnover scale appropriate for eddies of size $r$. Its global and local values are defined according to
\begin{equation}
\tau_{e} = L^{2/3} \langle \epsilon\rangle^{-1/3}, ~~~~\tau_{e,r} = r^{2/3} \epsilon_r^{-1/3} .
\end{equation}

\begin{figure}[htbp]
   \begin{center}
      \includegraphics[width=4.5 in] {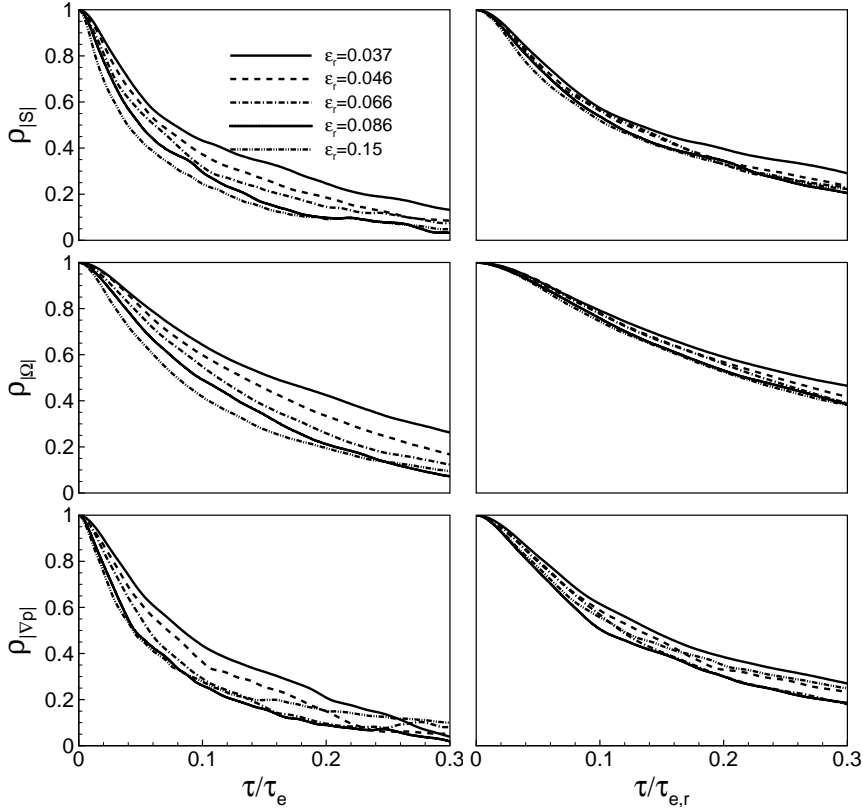}
      \caption{Auto-correlations of $|S|$ (top row), $|\Omega |$ (middle row), and $|\nabla p|$ (bottom row) vs. $\tau/\tau_e$ (left column) and $\tau/\tau_{e,r}$ (right column) for the case of 136$\eta_K$-II in Table \ref{ta:localcases}. }
   \label{Fig_collapse_64}
   \end{center}
\end{figure}

Fig. \ref{Fig_collapse_64} shows the conditional Lagrangian time correlations of absolute values of ${\bf S}$ (top row), ${\bf \Omega}$ (middle row), and $\bigtriangledown p$ (bottom row) with time normalized by $\tau_e$ (left column) and $\tau_{e,r}$ (right column). These results are for a single length-scale corresponding to the case of $r=136\eta_K$ (case II) in Table \ref{ta:localcases}. When the time lag $\tau$ is scaled by $\tau_e$, there are noticeable differences in the results depending on $\epsilon_r$. For the three variables, larger values of $\epsilon_r$ (i.e. in regions of more intense turbulence activity corresponding to smaller local eddy turn-over time) are associated with faster correlation decay. When the time lag is scaled by the local time-scale $\tau_{e,r}$, the curves  collapse better than with the global value. This provides some evidence for a Lagrangian RKSH also at inertial-range `eddy-turnover' scales.

\begin{figure}[htbp]
   \begin{center}
      \includegraphics[width=4.5 in] {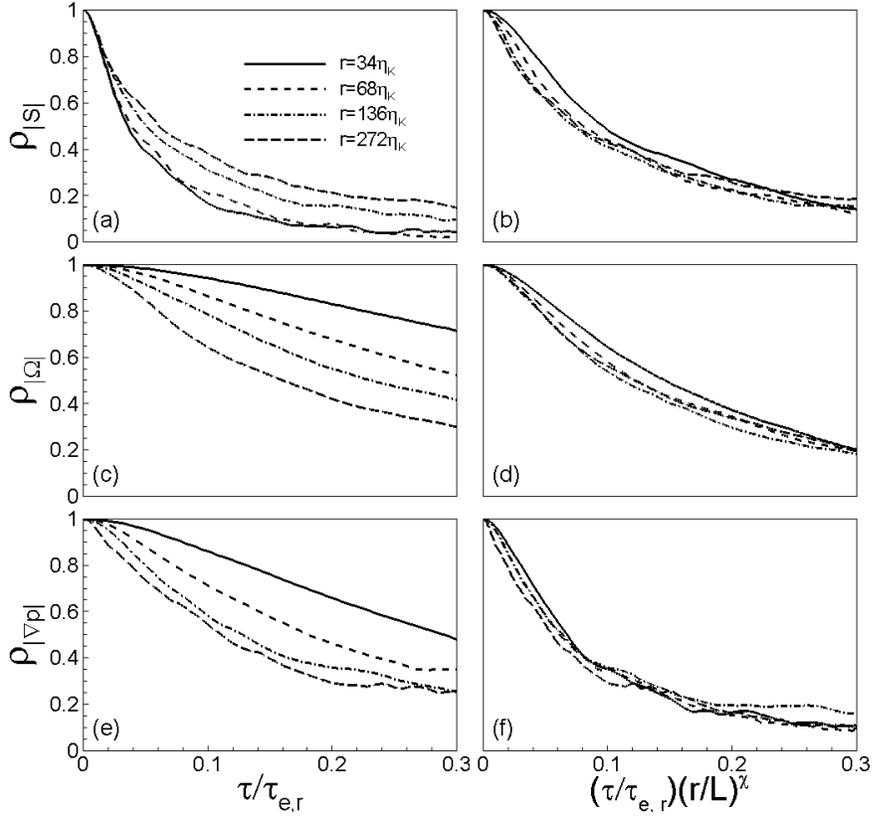}
      \caption{Time correlations of $|S|$, $|\Omega|$, and $|\nabla p|$ for the $\epsilon_r$ value in bin No. 3 (see Table \ref{ta:localcases}) vs. $\tau/\tau_{e,r}$ (left column) and $\tau/\tau_{e,r}\cdot \big(\frac{r}{L}\big)^{\chi}$ (right column, (b):$\chi=-0.3$ , (d):$\chi=-0.49$, and (f):$\chi=-0.55$). All are for for the cases in bin No. 3  with  $\epsilon\approx 0.092$, which also close to value of global-averaged dissipation-rate.}
      \label{Fig_collapse_er092}
   \end{center}
\end{figure}

The next question is whether good collapse also occurs for different length scales. Fig. \ref{Fig_collapse_er092} plots the conditional auto-correlations of $|S|$, $|\Omega|$, and $|\nabla p|$ with approximately the same $\epsilon_r$ corresponding to bin No. 3 (see Table \ref{ta:localcases}), but at different length scales. In the left column, the time lag is scaled by the local eddy time $\tau_{e,r}$. It is seen clearly that the correlation functions decay differently at different length scales, which implies that the normalization of time with $\tau_{e,r}$ does not account for the differences.

Intermittency in turbulence is often known to connect the inertial range dynamics with the ratio of length-scale to the integral scale, i.e. the level of intermittency is related to ``how far'' the scale is from its original starting point at the large scales during the cascade. Often such effects are parameterized by factors of the form $(r/L)^{\chi}$, where ${\chi}$ is an appropriate intermittency exponent for the correction. We determine the exponents ${\chi}$ empirically (see below) to obtain improved collapse. The right column in  Fig. \ref{Fig_collapse_er092} shows the correlation functions with time now normalized by the intermittency corrected time scale  $\tau_{e,r}(r/L)^{\chi}$. As can be observed, improved collapse can thus be obtained by using an intermittency correction.

\begin{figure}[htbp]
   \begin{center}
      \includegraphics[width=4. in] {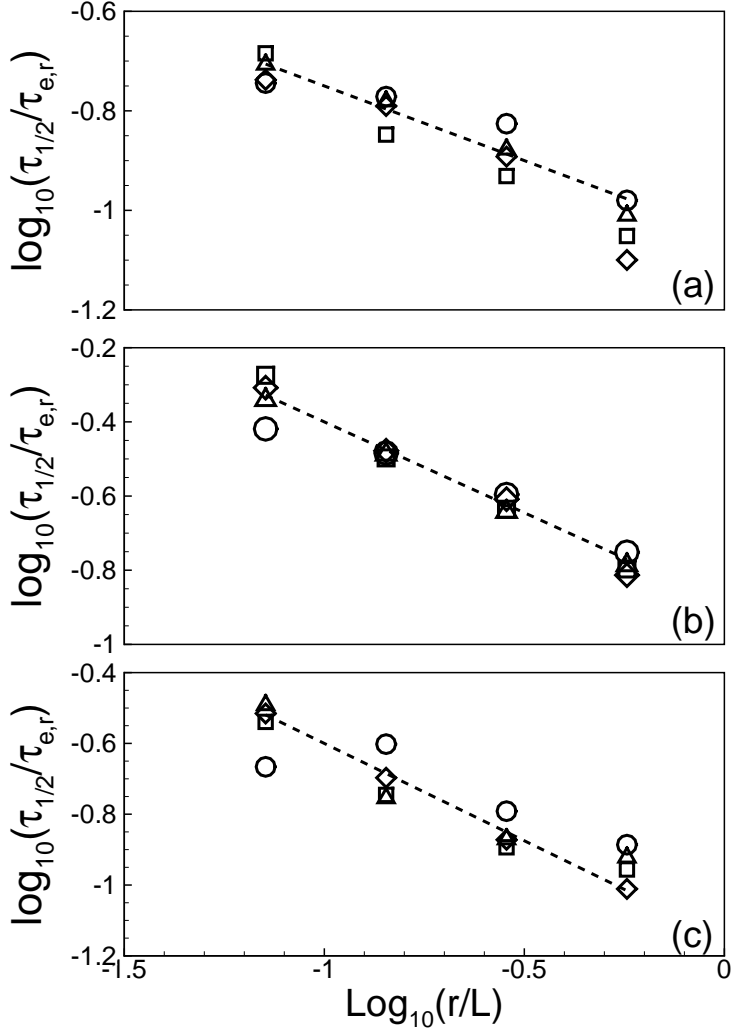}
      \caption{Plots of characteristic decay time $\tau_{1/2}$ (measured as the 1/2 point in the correlation function) versus length-scale.  The scale $r$ is  normalized by the integral length $L=1.367$. Different symbols are for different $\epsilon_r$ bins:  $\circ$: $bin_1$; $\bigtriangleup$: $bin_2$; $\Box$: $bin_3$, $\diamond$: $bin_4$. The dashed lines are power-law fits yielding, in (a) for $|S|$, $\chi=-0.3$; in (b) for $|\Omega|$, $\chi=-0.49$; and in (c) for $|\nabla p|$, $\chi=-0.55$.}
      \label{Fig_exponent}
   \end{center}
\end{figure}

In order to determine the exponents empirically, we set a threshold on the correlation function value. A value $\rho_f(\tau_{1/2}) = 1/2$ is used, which defines the time-scale $\tau_{1/2}$. For each case, the value of $\tau_{1/2}$ correspond to  $\rho_f = 1/2$. A log-log plot of $\tau_{1/2}/\tau_{e,r}$ versus $r/L$ should have slope $\chi$, if a power-law intermittency correction is appropriate. In Fig.  \ref{Fig_exponent} such plots are presented for each variable $|{\bf S}|$, $|{\bf \Omega}|$, $|{\bf \nabla p}|$,  from top to bottom. Exponents are obtained by fitting straight lines through the data as shown by the dashed lines in the plots. The corresponding exponents of $\chi$ are $\chi_{|S|} = -0.3$, $\chi_{|\Omega|} = -0.49$, and $\chi_{|\nabla p|} = -0.55$. These are the values used to scale the results shown in right column of Fig. \ref{Fig_collapse_er092}.
We have not yet succeeded in relating the values of ${\chi}$ to the multifractal theory of turbulence. Note that these values are significantly larger than what is typically obtained from multifractal corrections.

\begin{figure}[htbp]
   \begin{center}
      \includegraphics[width=4.5 in] {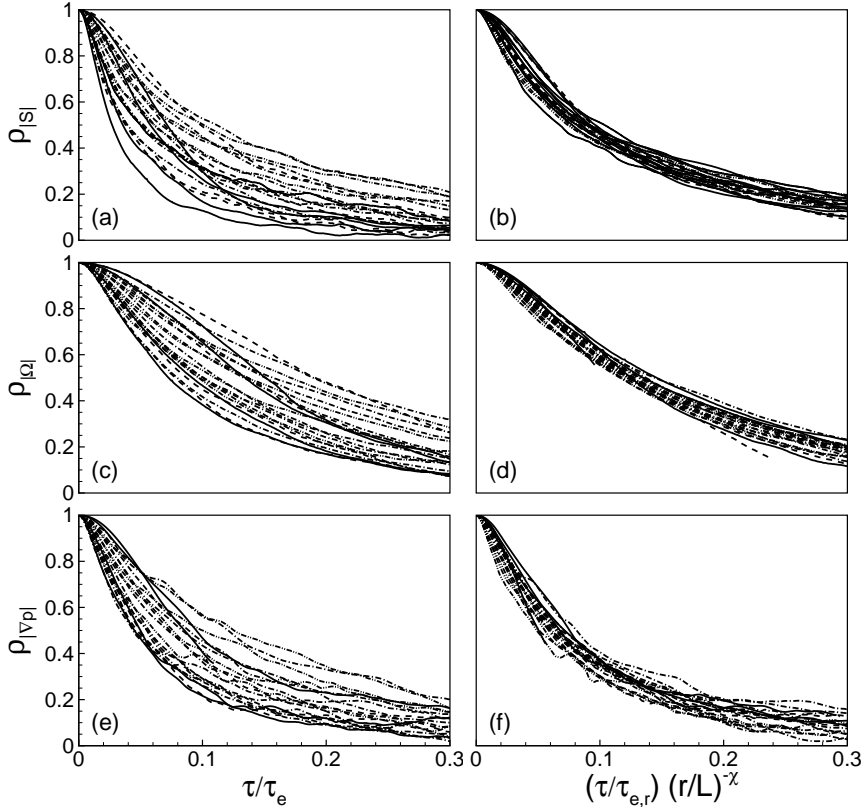}
      \caption{Time correlations of $|S|$ (top row), $|\Omega|$ (middle row), and $|\nabla p|$ (bottom row) vs. $\tau/\tau_e$(left column) and $\tau/\tau_{e,r}\cdot \big(\frac{r}{L}\big)^{\chi}$ (with (b):$\chi=-0.3$, (d):$\chi=-0.49$, and (f):$\chi=-0.55$) for all the cases in Table \ref{ta:localcases}. The lines are solid lines: $16$-cube; $--$ : $32$-cube; $-- \cdot$: $64$-cube; $-- \cdot \cdot$ : $128$-cube.}
      \label{Fig_collapse_all}
   \end{center}
\end{figure}

All the cases are plotted jointly in Fig. \ref{Fig_collapse_all}. With global time scaling of $\tau_e$ (left column), the curves scatter significantly reflecting clear intermittency in the flow. When both $\tau_{e,r}$ and $(r/L)^{\chi}$ are considered in the  scaling  of time lag (right column),  the curves collapse reasonably well. Some scatter remains, however.

\section{Conclusions}
\label{sec:concl}
Using the JHU public turbulence database, we perform Lagrangian analysis of temporal time correlations for the absolute values of velocity and pressure gradients.  Consistent with earlier results for the square strain-rate and rotation rates\cite{rf:Guala07,rf:Yeung07}, we find significantly longer decay times for the magnitudes as compared to the tensor or vector elements, especially for strain-rate and pressure gradients. It is demonstrated that Lagrangian dynamics of velocity gradient and pressure gradient (almost equivalent to acceleration since viscous effects are negligible) is determined mainly by scales provided by the locally averaged rate of dissipation, as predicated in the Kolmogorov Refined Similarity Hypothesis.  We point out that a Lagrangian KRSH has also been shown to hold in the context of moments of two-time velocity increments\cite{rf:Benzi08}. That analysis is based on a fully Lagrangian rate of dissipation  $\epsilon_{\tau}$ averaged over {\it temporal} domains of duration $\tau$ along the particle trajectory.  By its nature, $\epsilon_{\tau}$  averages dissipation at various times. The present analysis is instead based on the more often used spatial average of dissipation at a single (initial-condition) time. Still, present results show that even after using local time-scales corrected for intermittency, there was remaining scatter observed in the correlation functions for absolute value variables. A better understanding of the origin of these deviations, as well as relating the relatively large intermittency corrections  to various phenomenological models of turbulence, would be desirable developments.


\begin{thebibliography}{}
\bibitem{rf:Yu09a}
 Yu, H. and Meneveau, C.: Lagrangian refined Kolmogorov similarity hypothesis for gradient time-evolution in turbulent flows, submitted to Phys. Rev. Lett., 2009.
\bibitem{rf:Taylor21}
  Taylor, G. I.: Diffusion by continuous movements, Proc. London Math. Soc. {\bf 20}, 196 (1921)
\bibitem{rf:Richardson26}
  Richardson, L. F.: Atmospheric diffusion shown on a distance-neighbor graph, Proc. Roy. Soc. London, Ser. A {\bf 110}, 709 (1926)
\bibitem{rf:Kolmogorov41}
 Kolmogorov, A. N.: The local structure of turbulence in incompressible
viscous fluid for very large Reynolds numbers, Dokl. Akad. Nauk SSSR {\bf 30}, 301 (1941); also Proc. R. Soc. A {\bf 434}, 9 (1991)
\bibitem{rf:Kolmogorov62}
 Kolmogorov, A. N.: A refinement of previous hypotheses concerning the local structure of turbulence in a viscous incompressible fluid at high
 Reynolds number, J. Fluid Mech. {\bf 13}, 82 (1962)
\bibitem{rf:Stolovitzky92}
 Stolovitzky, G., Kailasnath, P., and Sreenivasan, K. R.: Kolmogorov's refined similarity hypotheses, Phys. Rev. Lett. {\bf 69}, 1178 (1992).
 \bibitem{rf:Thoroddsen}
 Thoroddsen, S. T., and van Atta, C. W.: Experimental evidence supporting Kolmogorov's refined similarity hypothesis, Phys. Fluids A {\bf 4}, 2592 (1992).
 \bibitem{rf:Chen93}
 Chen, S., Doolen, G. D., Kraichnan, R. H., and She, Z.-S., On statistical correlations between velocity invrements and locally averaged dissipatopn in homogenous turbulence, Phys. Fluids A {\bf 5} 458 (1993).
 \bibitem{rf:Stolovitzky94}
 Stolovitzky, G. and Sreenivasan, K. R.: Kolmogorov's refined similarity hypotheses for turbulence and general stochastic processes, Rev. Mod. Phys. {\bf 66}, 229 (1994).
 \bibitem{rf:Chen95}
 Chen, S., Doolen, G. D., Kraichnan, R. H., and Wang, L.-P.: Is the Kolmogorov refiend simmilarity relation dynamic or kinematic,  Phys. Rev. Lett. 74, 1775(1995).
 \bibitem{rf:Ching08}
 Ching, E. S. C., Guo, H., and  Lo, T. S.: Lagrangian properties of particles in turbulence, Phys. Rev. E {\bf 78}, 026303 (2008).
 \bibitem{rf:Yeung06}
 Yeung, P. K., Pope, S. B., Lamorgese, A. G., and Donzis, D. A.: Acceleration and dissipation statistics of numerically simulated isotropic turbulence, Phys. Fluids {\bf 18}, 065103 (2006).
 \bibitem{rf:Pope94}
  Pope, S. B.: Lagrangian PDF methods for turbulent flows, Annu. Rev. Fluid Mech.{\bf 26}, 23 (1994).
\bibitem{rf:Yeung02}
 Yeung, P. K.: Lagrangian investigations oF turbulence, Annu. Rev. Fluid Mech. {\bf 34},115 (2002).
\bibitem{rf:Pope85}
 Pope, S.B.: PDF methods for turbulent reactive flows, Prog. Energy Combust. Sci. {\bf 11}, 119 (1985).
\bibitem{rf:Girimaji90a}
 Girimaji, S.S. and Pope, S.B.: Material element deformation in isotropic turbulence, J Fluid Mech. {\bf 220}, 427 (1990).
 \bibitem{rf:Girimaji90}
  Girimaji, S. S. and Pope, S. B.: A diffusion model for velocity gradients in turbulence, Phys. Fluids A {\bf 2}, 242 (1990).
 \bibitem{rf:Vieillefosse82}
   Vieillefosse, P.: Local interaction between vorticity and shear in a perfect incompressible fluid, J. Phys. (France) {\bf 43}, 837 (1982).
\bibitem{rf:Cantwell92}
  Cantwell, B. J.:  Exact solution of a restricted Euler equation, Phys. Fluids A {\bf 4}, 782(1992).
\bibitem{rf:Martin98}
 Martin, J., Ooi, A., Chong, M. S., and Soria, J.: Dynamics of the velocity gradient tensor invariants in isotropic turbulence, Phys. Fluids {\bf 10}, 2336 (1998).
\bibitem{rf:Chertkov99}
  Chertkov, M., Pumir, A., and Shraiman, B. I.: Lagrangian tetrad dynamics and the phenomenology of turbulence, Phys. Fluids {\bf 11}, 2394 (1999).
\bibitem{rf:Jeong03}
  Jeong  E., and Girimaji, S. S.: Velocity-gradient dynamics in turbulence:Effect of viscosity and forcing, Theor. Comput. Fluid Dyn. {\bf 16}, 421 (2003).
\bibitem{rf:Chevillard06}
  Chevillard, L., and Meneveau, C.: Lagrangian dynamics and statistical geometric structure of turbulence, Phys. Rev. Lett. {\bf 97}, 174501(2006).
\bibitem{rf:Biferale07}
  Biferale, L., Chevillard, L., Meneveau, C.  and Toschi, F.: Multiscale model of gradient evolution in turbulent flows, Phys. Rev. Lett. {\bf 98}, 214501 (2007).
\bibitem{rf:Guala07}
 Guala, M., Liberzon, A., Tsinober, A., and Kinzelbach, W.: An experimental investigation on Lagrangian correlations of small-scale turbulence at low Reynolds number, J. Fluid Mech. {\bf 574}, 405 (2007).
\bibitem{rf:Yeung07}
 Yeung, P. K., Pope, S. B., Kurth, E. A., and Lamorgese, A. G.: Lagrangian conditional statistics, acceleration and local relative motion in numerically simulated isotropic turbulence, J. Fluid Mech. {\bf 582}, 399 (2007).
 \bibitem{rf:Riley74}
  Riley, J. J. and Patterson, G. S.: Diffusion experiments with numerically integrated isotopic turbulence, Phys. Fluids {\bf 17}, 292 (1974).
\bibitem{rf:She91}
 She, Z-S., Jackson, E., and Sreenivasan, K. R.: Structure and dynamics of homogenous turbulence: models and simulations, Proc. R. Soc. Lord. A {\bf 434}, 101 (1991).
\bibitem{rf:Toschi09}
  Toschi F. and Bodenschatz, E.: Lagrangian properties of particles in turbulence, Annu. Rev. Fluid Mech.{\bf 41},375 (2009).
 \bibitem{rf:Li08}
 Li, Y.,  Perlman, E.,  Wan, M., Yang, Y., Burns, R., Meneveau, C., Burns, R., Chen, S.,  Szalay, A.,  and Eyink, G.: A public turbulence database cluster and applications to study Lagrangian evolution of velocity increments in turbulence, J. Turbulence {\bf 9}, 31 (2008).
\bibitem{rf:Yeung88}
  Yeung, P. K. and Pope, S. B.: An algorithm for tracking fluid particles in numerical Simulations of homogeneous turbulence, J. Comp. Phys. {\bf 79}, 373 (1988).
\bibitem{rf:Yu09b}
  Yu H., and Meneveau, C.: Lagrangian time-correlations of strain and rotation rates in isotropic turbulence, in preparation to be submitted to Phys. Fluids, 2009.
\bibitem{rf:Benzi08}
 Benzi, R., Biferale, L., Calzavarini, E., Lohse, D., and Toschi, F.: Velocity gradients along particles trajectories in turbulent flows, arXiv:0806.4762v1 [physics.flu-dyn] 30 Jun 2008
\end{thebibliography}
\end{document}